\documentclass[12pt,preprint]{aastex}

\shorttitle{Cygnus Loop Soft X-rays}
\shortauthors{McEntaffer et al.}

\begin{document}

\title{Soft X-ray Spectroscopy of the Cygnus Loop Supernova Remnant}

\author{R. L. McEntaffer and W. Cash}
\affil{Center for Astrophysics and Space Astronomy, University of Colorado, Boulder, CO 80309}
\email{randall.mcentaffer@colorado.edu}

\begin{abstract}
The Cygnus X-ray Emission Spectroscopic Survey (CyXESS) sounding rocket payload was launched from White Sands Missile Range on 2006 November 20 and obtained a high resolution spectrum of the Cygnus Loop supernova remnant in the soft X-rays.  The novel X-ray spectrograph incorporated a wire-grid collimator feeding an array of gratings in the extreme off-plane mount which ultimately dispersed the spectrum onto Gaseous Electron Multiplier (GEM) detectors.  This instrument recorded 65 seconds of usable data between 43-49.5 \AA\ in two prominent features.  The first feature near 45 \AA\ is dominated by the He-like triplet of \ion{O}{7} in second order with contributions from \ion{Mg}{10} and \ion{Si}{9}-\ion{Si}{12} in first order, while the second feature near 47.5 \AA\ is first order \ion{S}{9} and \ion{S}{10}.  Fits to the spectra give an equilibrium plasma at $\log(T)=6.2$ ($kT_e=0.14$ keV) and near cosmic abundances.  This is consistent with previous observations, which demonstrated that the soft x-ray emission from the Cygnus Loop is dominated by interactions between the initial blast wave with the walls of a precursor formed cavity surrounding the Cygnus Loop and that this interaction can be described using equilibrium conditions.
\end{abstract}



\section{Introduction}\label{intro}
Supernovae greatly influence the dynamics within the interstellar medium (ISM).  Their ubiquitous nature and the size of their remnants allow them to influence multiple phases of the ISM while influencing the evolution and structure of the galaxy.  They are responsible for chemical enrichment of the ISM, play a major role in energy input, and are dominant sources for hot ionized gas, which fills much of the galaxy \citep{McKeeOstr77}.

One of the most studied SNRs is the Cygnus Loop.  It is a nearby remnant at 540 pc \citep{blair05}, quite large, filling $\sim3^{\circ}\times3^{\circ}$ \citep{L97}, and is relatively unabsorbed ($E(B-V)=0.08$; \cite{Fesen}).  The blast wave is currently encountering the surrounding ISM leading to bright emission at all wavelengths.  Therefore, the Cygnus Loop serves as an excellent test bed for SNR evolution theory.  A comprehensive and detailed imaging study of the remnant's X-ray and optical emission was performed by \citet{L97,L98}.  One of the major findings of these studies is that the Cygnus Loop morphology is not indicative of the typical theoretical picture of a blast wave propagating through a uniform medium \citep{sedov,spitzer}, but instead is an interaction of the blast wave with an inhomogeneous medium.  The precursor formed cavity is nearly spherical and surrounded by high density clumps and lower density gas.  This causes most emission to originate from a limb-brightened shell.  The optical emission occurs when the shock is decelerated rapidly in high density clumps.  The X-ray emission occurs in these clumps as the blast wave interacts with the cloud and hardens toward the interior as a reflection shock propagates inward reschocking the already shocked gas.  X-rays also originate as expected in the lower density gas that follows the theoretical model more closely.

The Cygnus Loop has been spectroscopically observed in the soft X-ray to some extent.  \citet{vedder} obtained a high resolution (E/$\Delta$E$\sim50$) spectrum at energies below 1 keV using the Focal Plane Crystal Spectrometer (FPCS) on the Einstein Observatory, which observed a $3' \times 30'$ bright northern section midway between the center and the limb.  They detected \ion{O}{8} Lyman-$\alpha$ at 653 eV, the resonance line in the He-like triplet of \ion{O}{7} at 574 eV and a blend of \ion{Ne}{9} lines between 905-920 eV.  They do not detect the forbidden or intercombination lines of the \ion{O}{7} triplet.  Even though the detected fluxes are consistent with an equilibrium plasma with solar abundances at $T\sim3\times10^6$ K, the authors argue that the plasma is not yet in collisional ionization equilibrium (CIE) due to the lack of forbidden \ion{O}{7} emission.

ROSAT Position Sensitive Proportional Counter (PSPC) observations by \citet{L99} and \citet{M01} show results consistent with the previous imaging studies.  A \citet{L99} spectrum of the decelerated cloud shock was best fit by a Raymond-Smith equilibrium plasma at cosmic abundances \citep{RaySmith} with $kT_e=0.06(+0.04, -0.02)$ keV and absorbing column $N_H=7\times10^{20}(\pm4\times10^{20})$ cm$^{-2}$.  A spectrum extracted from a region interior to the cloud shock had a similar low temperature component, but also included a high temperature contribution, $kT_{e1}=0.25(+0.14, -0.06)$ keV, from the reflected shock, which hardens the interior emission.  The low temperature corresponds to a shock velocity of $v_s=230$ km s$^{-1}$, which has slowed considerably from the initial blast wave velocity of $v_s\sim500$ km s$^{-1}$ \citep{L98} resulting in an equilibrium plasma.  \citet{M01} also find that equilibrium plasmas with cosmic abundances fit the low temperature components of their extracted spectra ($kT_e=0.043$ to $0.067$ keV) and that these contributions are greatest toward the limb.  However, they use ASCA data and a nonequilibrium model with depleted abundances to fit the interior harder spectra ($kT_e=0.27$ to $0.34$ keV).

The Chandra X-ray Observatory Advanced CCD Imaging Spectrometer (ACIS) also observed the Cygnus Loop \citep{L02,leahy}.  \citet{L02} used the Xspec \citep{Xspec} spectral fitting software package and more specifically the MEKAL equilibrium model \citep{Mewe85,Mewe86,kaastra,arnaud85,arnaud92,liedahl95} to fit the extracted spectra.  Again, the softer emission originates near the limb ($kT_e\sim0.03$ keV) while the reflected shock hardens the interior spectrum ($kT_e=0.12$ to $kT_e=0.18$ keV).  The best spectral fits all required a depletion in oxygen.  Attempts were made at nonequilibrium models but they did not improve the fit statistics.  Furthermore, the fit values for the ionization parameter ($n_et$, where $n_e$ is electron density and $t$ is the elapsed time since the gas was initially shocked) were all $>10^{12}$ cm$^{-3}$ s where values of $\gtrsim3\times10^{11}$ cm$^{-3}$ s signify equilibrium.  The authors do not rule out the possibility of nonequilibrium but state that due to low spectral and spatial resolution the low temperature equilibrium plasma is indistinguishable from a higher temperature nonequilibrium plasma.  Using 21 different extraction regions \citet{leahy} finds the same result; the spectra are best fit by an equilibrium MEKAL model with variable elemental abundances, and that there is no evidence for nonequilibrium.  The author also states that the abundances are considerably depleted and vary on small spatial scales suggesting that the region is geometrically complex with multiple clouds and even more shocks.

The final spectrum of note \citep{M07} was taken using the Suzaku Observatory X-ray Imaging Spectrometer (XIS) CCD camera and has the highest spectral resolution other than \citet{vedder}.   The authors clearly detected \ion{O}{7} at $562\pm10$ eV, \ion{O}{8} at $653\pm10$ eV, \ion{C}{6} at $357\pm10$ eV and \ion{N}{6} at $425\pm10$ eV, but the 1/4 keV band emission is unresolved.  Band ratio maps using strong emission features at different radii show that the outermost emission is dominated by the 1/4 keV emission and therefore a low temperature component.  Furthermore, the \ion{O}{8}/\ion{O}{7} ratio increases inward showing that the ionization state is higher in the interior.  The best fit models were two component nonequilibrium plasmas with variable abundances (\textit{vnei} in Xspec).  The lower temperature components of the fits varied from $kT_e=0.10-0.15$ keV and the high temperature component ranged from $kT_e=0.18-0.34$ keV.  The temperature of both components increased towards the interior.  Column densities ranged from $N_H=3-6\times10^{20}$ cm$^{-2}$ and abundances were heavily depleted.

The Cygnus Loop is clearly complex both spatially and spectrally.  However, spectral resolution is lagging spatial resolution.  Chandra can image fine structures indicative of different physical regions, but the nearly broadband spectra can only show trends.  Determining plasma diagnostics from model fits to these low spectral quality data is uncertain.  Higher spectral resolution is required not only to constrain the parameters of a model, but to test the assumed validity of the model.

Currently there is no efficient, well developed technology that permits high resolution x-ray spectroscopy from large solid angle sources, making spectra of diffuse x-ray sources rare.  In order to address this issue a soft X-ray spectrometer was designed for a rocket payload, the Cygnus X-ray Emission Spectroscopic Survey (CyXESS).  Scientifically, the payload was designed to observe emission in the 1/4 keV bandpass.  This is the least understood range of astrophysical soft X-ray energies.  A diffuse high resolution spectrum has never been achieved even though there is a large amount of flux in this band.  The soft emission from the Cygnus Loop is contained within a narrow shell and is dominated by the early interactions of the initial blast wave with the surrounding cavity.  Therefore, a global spectrum of the remnant will be dominated by the physics of this interaction.

\section{Sounding Rocket Instrument}\label{inst}
Typical X-ray telescopes employ the use of grazing incidence telescopes.  However, these are expensive and heavy and thus unattractive for a rocket flight.  CyXESS utilizes a wire grid collimator to constrain the beam of light.  The collimator is followed by the off-plane reflection grating array which disperses light onto the detectors.  The payload is three meters long consisting of nearly a meter to create the converging beam while allowing the gratings to throw the light about two meters.  A brief description of the payload design follows, but a detailed description can be found in \citet{McE07,McE08}.

\subsection{Wire Grid Collimator}\label{coll}
Wire grid collimators have wires that are spaced periodically in such a way that only light coming from a specified direction can pass through.  If the grids have a spacing that decreases systematically, then it is possible to allow only light that is converging to a line to pass through, simulating the output of a lens.  As light travels from front to back in the collimator it will encounter the same number of slits but they will be narrower and closer together, thus sculpting the converging beam.  The wires create baffles between slits which vignette unwanted rays.  If thin material is used, these wire spacers serve as knife edges so that any light striking the metal will be near normal incidence and will be efficiently absorbed.  Such a collimator does not function well for a point source, but for a diffuse target, radiation comes from all directions, and the beam is fully illuminated.  A grating mounted in the exit beam diffracts just as in a telescope beam.  Thus the collimator alone provides the needed beam geometry.

\subsection{Off-plane Grating Array}\label{grats}
The off-plane mount at grazing incidence brings light onto the grating at a low graze angle, quasi-parallel to the direction of the grooves.  The light is then diffracted through an arc, forming a cone, so that this mount is also known as conical diffraction \citep{cash91,catura}.  The off-plane grating equation is
\begin{equation}
\sin\alpha+\sin\beta=\frac{n\lambda}{d\sin\gamma},
\label{grat}
\end{equation} \noindent
where $d$ is the spacing between grooves and $\gamma$ is the angle between the direction of the incoming ray and the direction of the groove at the point of impact.  Light comes into the grating at an azimuthal angle of $\alpha$ along a cone with half-angle $\gamma$.  It is then diffracted along the same cone of half-angle $\gamma$, but now with an azimuthal angle of $\beta$.  If $\gamma$ is kept small then the arc of diffraction stays close to the plane of the grating.  However, at these low angles even large gratings have a small cross section to the incoming light.  The solution is to place gratings in an array.

The flight gratings have sinusoidal grooves with a density of $5670\pm10$ grooves/mm and an active area of 104 mm $\times$ 20 mm with the latter being the dimension along the groove.  Using the limitations set by module packaging inside the payload as well as detector positioning on the bulkhead, the optimal graze angle is 4.4$^{\circ}$.  A 20 mm grating at 4.4$^{\circ}$  gives only a 1.53 mm effective length.  Therefore, in order to maximize the throughput of the grating array, the grating substrates are extremely thin to minimize the amount of light they block.  We used electroformed nickel with a figure of $\sim\lambda$/2 and thickness of $0.127\pm0.008$ mm.  After accounting for the thickness, 67 gratings per module are required to catch the light in a 110 mm $\times$ 110 mm array.  The array was held in a tension mount ($\sim22$ N per grating) to maintain the figure on the gratings and ensure survivability during launch.

\subsection{GEM Detectors}\label{gems}
The two detectors on the payload are Gaseous Electron Multipliers (GEM).  They were built by Sensor Sciences, LLC.  These innovative detectors use a gas filled chamber (75\% Ar 25\% CO$_2$) segmented by perforated polyimide (Kapton\textregistered) film coated with a conductive layer on each side.  The perforation holes provide a potential difference through which the electron cloud is accelerated resulting in gain.  One of the most attractive features of these detectors is that they are made with very large formats, which is essential for this experiment due to the system's dispersion and line lengths.  The entrance window is a 105 mm $\times$ 105 mm polyimide window that is 3600-3900 \AA\ thick to maximize transmission while maintaining integrity.  A 100 \AA\ carbon coat was added for conductivity and a grid bar and mesh support system is utilized with a transmission of 57.8\%.  The mesh and grid bars carry the negative high volts (HV) so that electrons are accelerated towards the anode, which is held at ground.  Gain is determined by many factors including the voltage drop across the foils and gaps, high voltage supply stability, cleanliness of the GEM foils, gas pressure, etc.  A gas flow system was incorporated to replenish the counter gas and maintain an operating pressure of 14.5 psia.  This system also counteracts the leak rate and compensates for micro tears in the window in order to improve gain stability.

The 100 mm $\times$ 100 mm anode is a serpentine cross delay line.  The output of the resistive anode is analyzed by a custom electronics system.  Signals are passed through an amplifier and then to an adder box which combines the data of the two detectors and passes it on to the timing-to-digital converter (TDC).  The TDC returns a 12-bit word for X position and Y position and an 8-bit word for pulse height.  The least significant bit of the pulse height determines which detector the current data word originates from.  Finally, a stim pulse is sent from the TDC to the anode, which is then analyzed and sent back to the TDC.  This gives reference data on both position and pulse height.  This stim can be seen without HV on and thus provides a useful diagnostic.

\subsection{Expected Performance}\label{expperf}
\begin{figure} [htbp]
   \centering
   \plotone{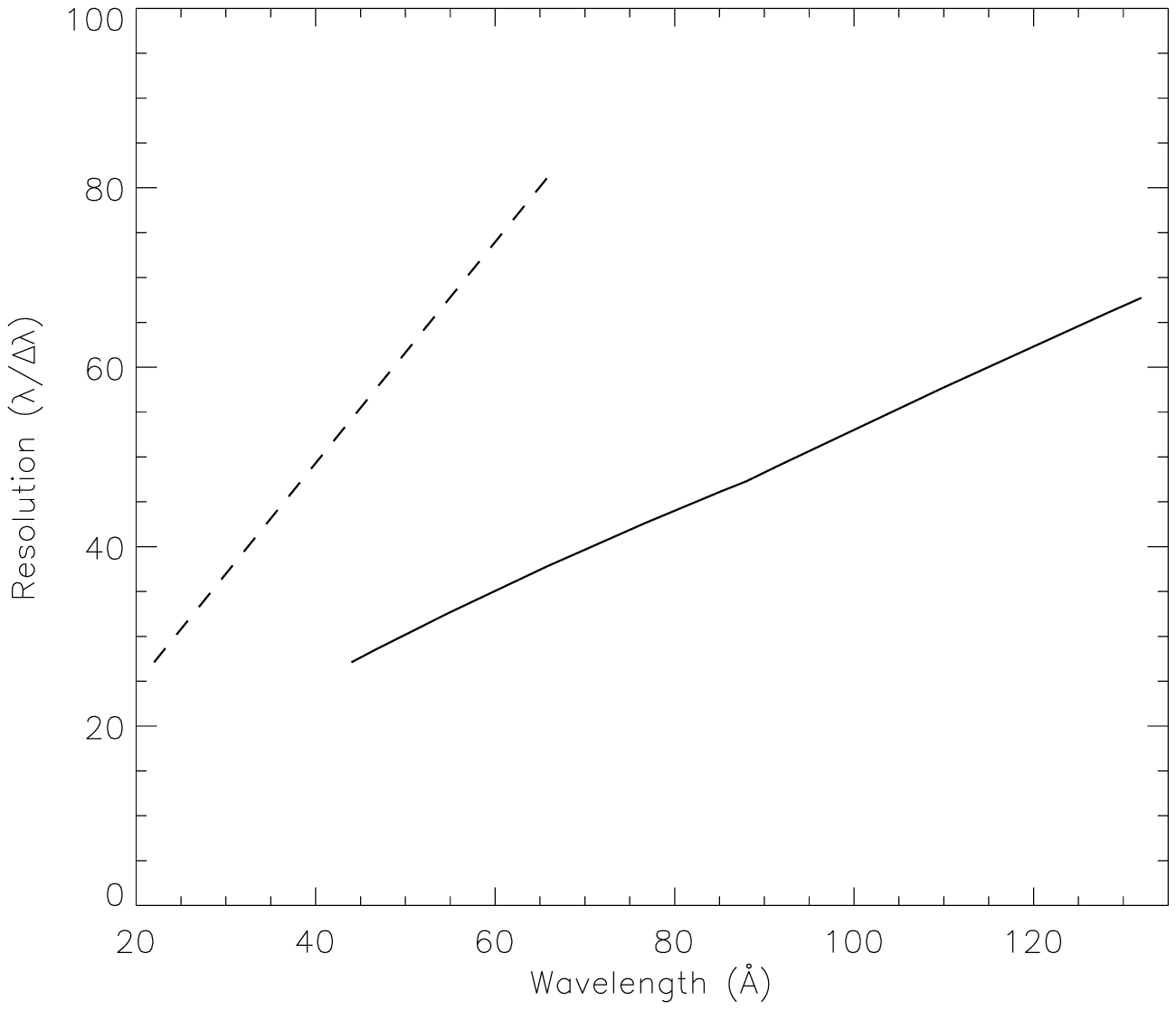}
   \caption{Spectral resolution as a function of wavelength for first order (solid line) and second order (dashed line).}
   \label{res}
\end{figure}

The payload was designed to obtain spectra from diffuse sources in the soft X-ray.  Several design factors determine the accepted passband such as the length of the payload, the size of the detectors and the dispersion of the gratings.  Optimizing these factors resulted in a passband of 44 \AA\ to 132 \AA\ in first order.  The expected performance at these wavelengths can be summarized by the resolution and effective area of the payload.  The resolution is ultimately determined by the full-width at half maximum (FWHM) of a spectral line at the focus along with the dispersion of the system.  The grating groove density of 5670 grooves/mm and throw distance of $\sim$2 m give a dispersion of 0.89 \AA\ mm$^{-1}$ in first order.  Calibrations of the spectrum give line widths that broaden slightly from 1.7 mm to 2.2 mm as wavelength is increased.  These characteristics give resolution of $\lambda/\Delta\lambda\sim25-70$ in first order and $\sim25-85$ in second order (22-66 \AA) as shown in figure~\ref{res}.

\begin{figure} [htbp]
   \centering
   \plotone{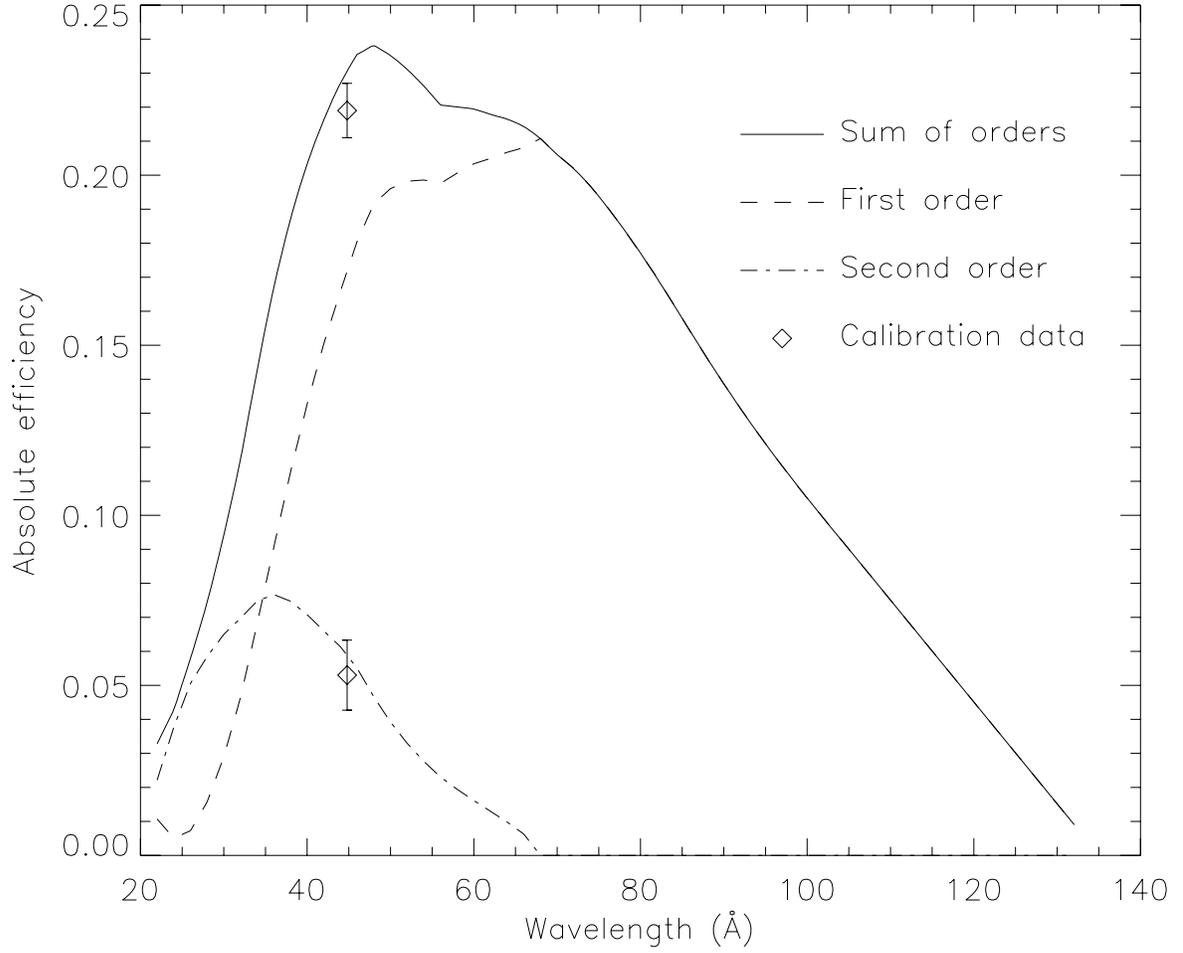}
   \caption{Theoretical grating efficiency curves for first order (dashed), second order (dot-dashed) and sum of orders (solid).  The data points are from pre-flight calibrations using carbon K-shell emission.}
   \label{efficcal}
\end{figure}
\begin{figure} [htbp]
   \centering
   \plotone{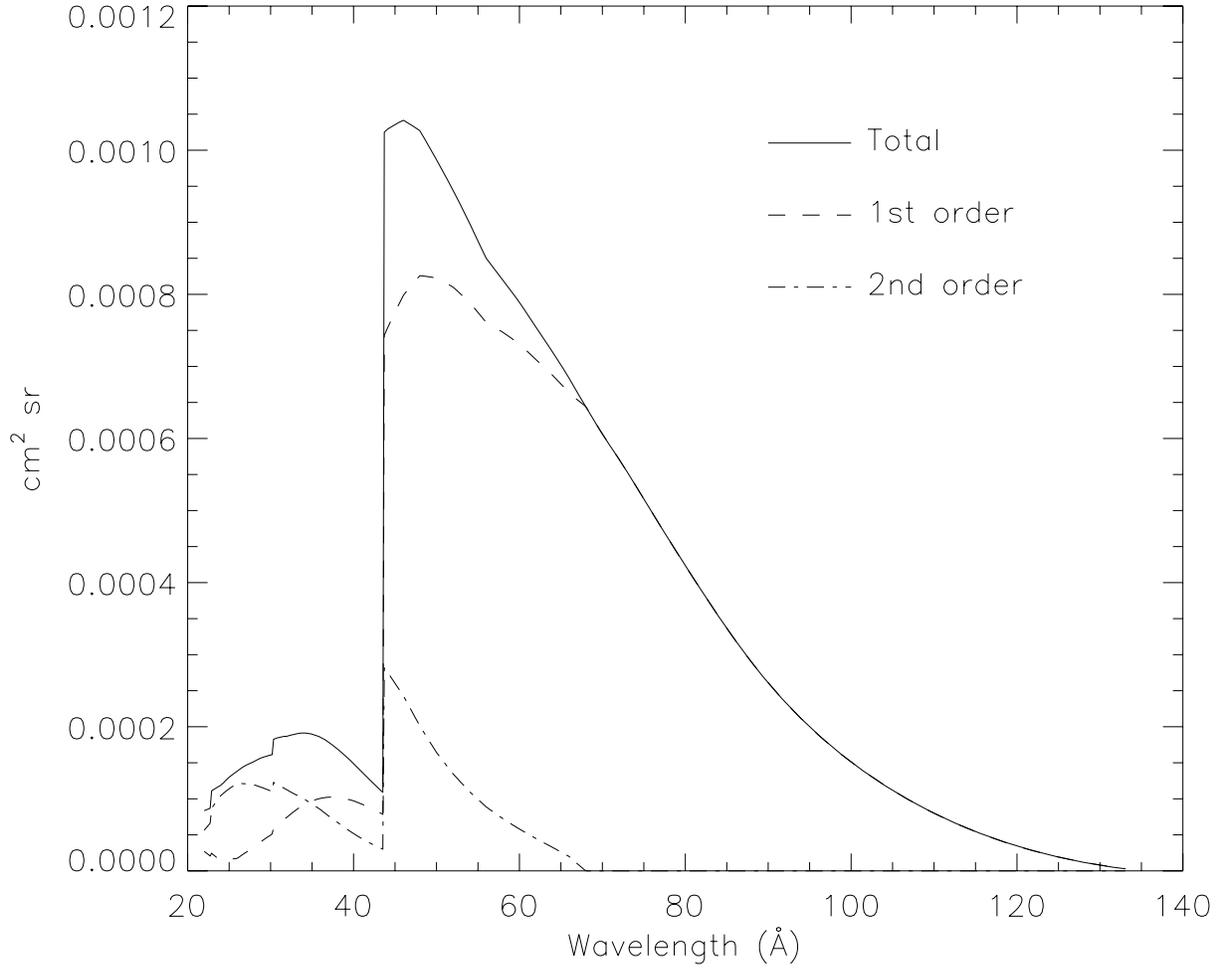}
   \caption{Effective area curves for first order (dashed), second order (dot-dashed) and sum of orders (solid).}
   \label{effarea}
\end{figure}

The effective area of this spectrograph is determined by the collecting area, sky coverage, and throughput of the system.  The collecting area of the telescope is defined by the size of the zero order image at the focal plane; 1.7 mm wide line over 10 cm of detector gives 1.7 cm$^2$.  This small amount of collecting area is bolstered by the large amount of solid angle available to each point on the focal plane, 8.93 deg$^2$.  In terms of efficiency, the detector gas absorbs all X-rays, but the mechanical throughput is 57.8\%, which is then further reduced by the transmission of the polyimide/carbon window.  As for the gratings, the theoretical efficiencies are plotted as lines in figure~\ref{efficcal}.  Calibration data are shown as the points at 44.76 \AA\ (carbon K-shell), with 1$\sigma$ Gaussian error bars, and agree well with theory.  Taking all factors into account, the resulting effective area curves are shown in figure~\ref{effarea}.



\subsection{Flight}\label{flight}
\begin{figure} [htbp]
   \centering
   \plotone{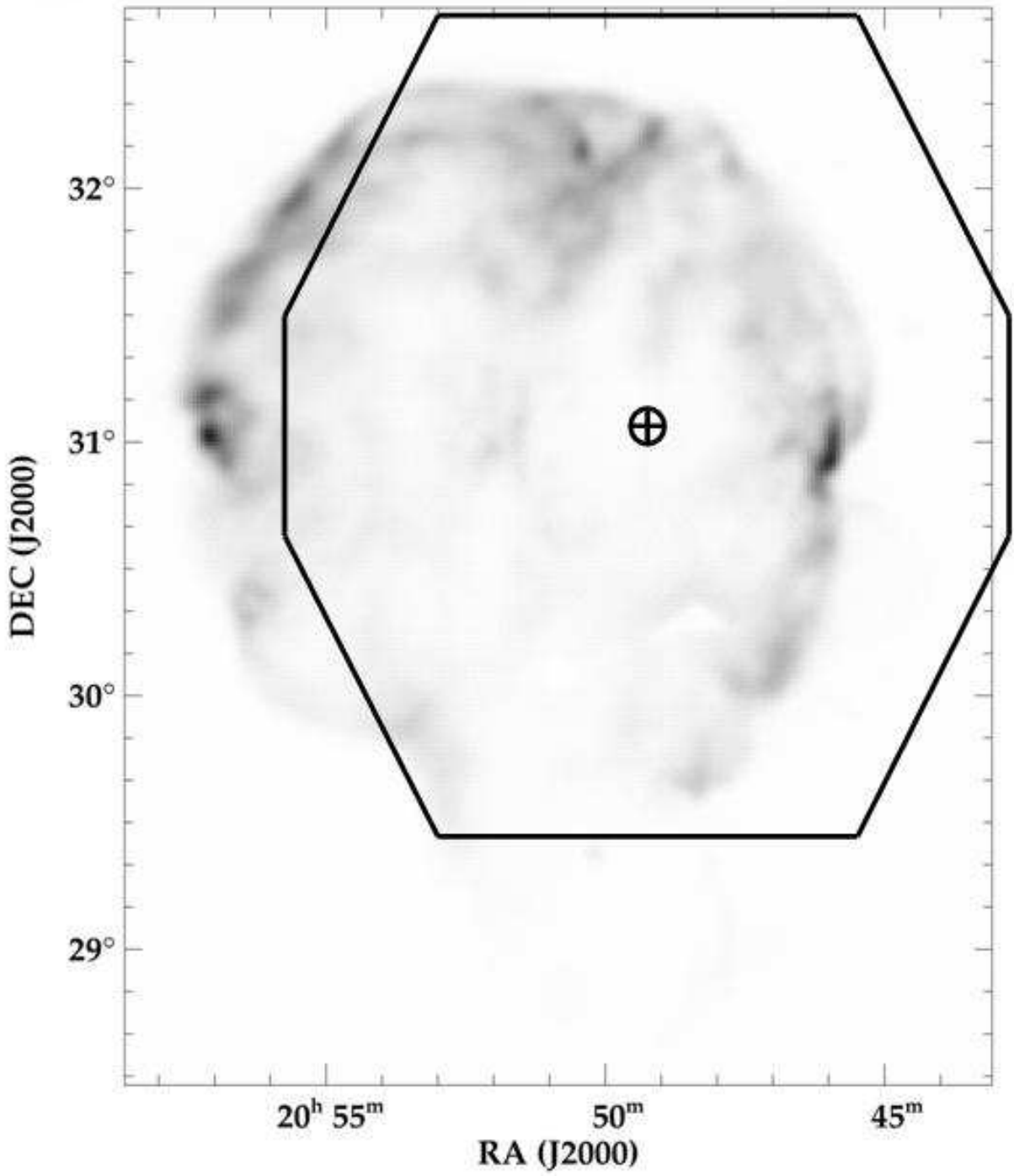}
   \caption{The octagonal FOV of CyXESS is shown over a ROSAT brightness map of the 0.25 keV band taken from \citet{L99}.  This pointing is $\sim55'$ to the west of the center position and is where the telescope was pointed when the spectral data were observed.}
   \label{finalfov}
\end{figure}

The payload was launched from White Sands Missile Range at 02:00:00 UT, 2006 November 21 (flight 36.224).  The zenith angle of the target was $\sim31^{\circ}$.  Usable data were recorded over 345 seconds of the flight.  However, a breakdown event upon high voltage turn on rendered one detector useless while leaving the other detector extremely noisy for most of the flight.  The GEM detectors exhibit noise in the form of hotspots which typically decay over time.  Therefore, usable spectral data were only recorded over 65 seconds near the end of the flight.  The pointing was dithered during the flight so that during the time when spectral data were collected the payload was pointed off center as depicted in figure~\ref{finalfov}.


\section{Data Analysis}\label{anal}
\begin{figure} [htbp]
   \centering
   \plotone{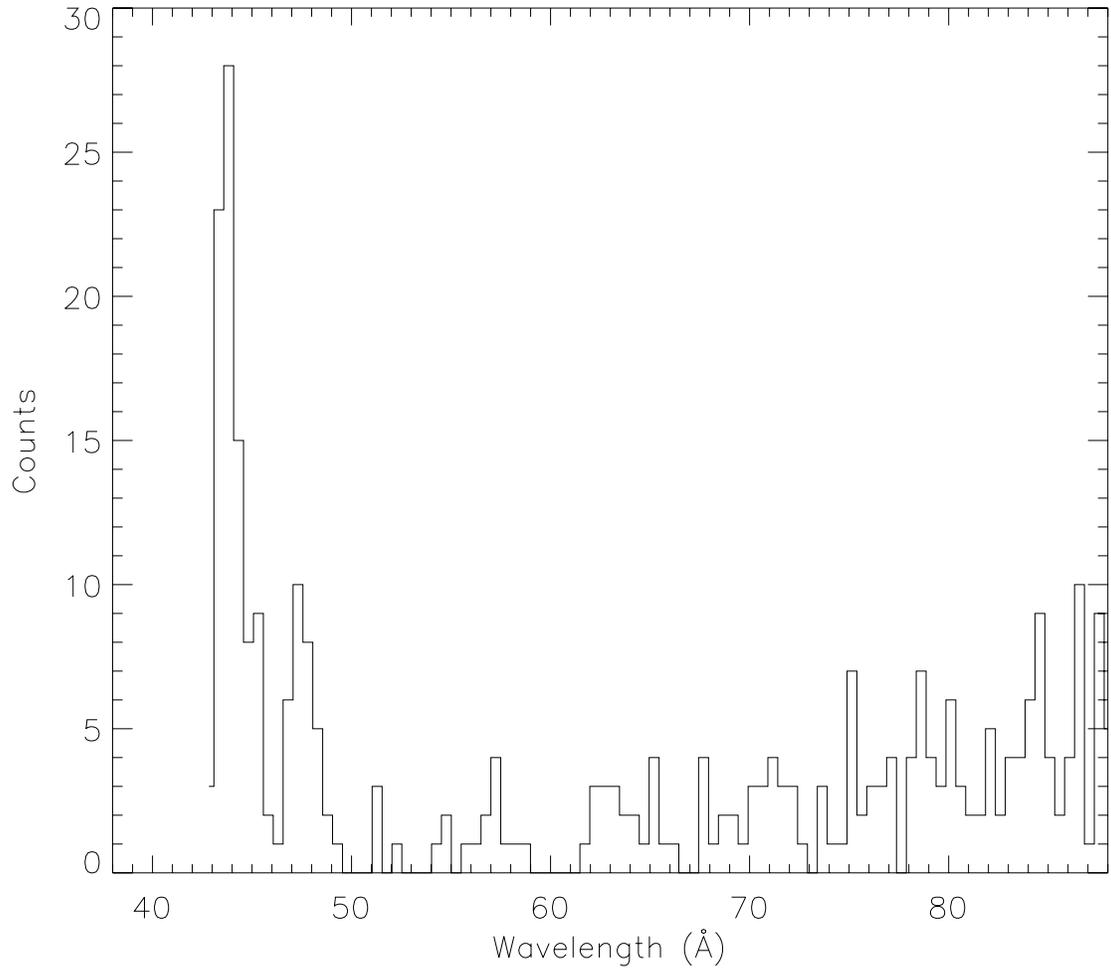}
   \caption{Two spectral lines stand out on the left side.  The rest of the counts are residual broad oval emission that dominates the noise.}
   \label{rawdata}
\end{figure}

Data were extracted from the section of the detector that was free of noise and contained spectral information.  The resulting spectrum is given in figure~\ref{rawdata}.  Pre-flight and post-flight spectral calibration data were compared with flight data to determine the wavelength scale.  The counts at wavelengths longer than 50 \AA\ are residual emission from a large hotspot that dominated the detector for most of the flight.  The two prominent features below 50 \AA\ are the detected spectral features.


To fit the data a series of equilibrium spectral models of an optically thin plasma under collisional ionization equilibrium were constructed using line lists from Raymond-Smith \citep{RaySmith}, MEKAL \citep{Mewe85,Mewe86,kaastra,arnaud85,arnaud92,liedahl95} and APED \citep{aped1,aped2} at temperatures ranging from $kT_e=0.034-0.272$ keV ($\log(T)=5.6-6.5$).  The spectra are constructed using Gaussians placed at the appropriate wavelengths (per Raymond-Smith, MEKAL or APED) with widths corresponding to the 1.84 mm (1.64 \AA) FWHM of the spectral line calibration data and amplitudes such that the integrated flux of each line scales according to the emissivities given in each line list.  Each model spectrum is absorbed using photoelectric absorption cross sections from \citet{morrison} with $N_H=7\times10^{20}$ cm$^{-2}$ (held constant) and then convolved with the payload instrument response function (figure~\ref{effarea}).  Cosmic abundances are per \citet{allen}: He, 10.93; C, 8.52; N, 7.96; O, 8.82; Ne, 7.92; Na, 6.25; Mg, 7.42; Al, 6.39; Si, 7.52; S, 7.20; Ar, 6.80; Ca, 6.30; Fe, 7.60; Ni, 6.30 (in logarithmic units where $\log_{10}N_H=12.00$).

Given the low number of counts, a maximum likelihood analysis was performed to obtain the best fit.  Assuming that the number of counts in each bin follows a Poisson distribution and that each bin is independent of the others results in the following likelihood function
\begin{equation}
\mathcal{L}=\displaystyle\prod_{i=1}^N\frac{Z_i^{n_i}exp(-Z_i)}{n_i!}.
\label{maxlikeli}
\end{equation}
The likelihood ($\mathcal{L}$) was numerically maximized to provide the best fit and the most likely values for the parameters of the model ($Z_i$) given the data ($n_i$).  The fit parameters include elemental abundances, which vary the amplitudes of corresponding lines, and a wavelength offset to the data, which adjusts the data wavelength scale within the errors of the detector calibration.  Several priors were applied to constrain the fit parameters.  First, abundances were assumed to always be $\geq 0$.  Second, the wavelength offset ($\lambda$ shift) was always contained within the limits set by detector calibrations.  Third, in all cases an attempt was made to vary as few abundances as possible while keeping abundances close to cosmic.  Finally, fits were made with the constraint that line fitting only occurred at wavelengths less than $\sim$50 \AA\ while model lines longward of this were depleted until they were less than 2$\sigma$ above the long wavelength noise.  The noise level was determined by a linear fit and the 2$\sigma$ upper limits were taken from \citet{gehrels} for each bin given the fit.  Only one model converged, a MEKAL plasma at $kT_e=0.14$ ($\log(T)=6.2$). The fit to the spectral data was obtained by varying only the S abundance and the wavelength offset.  The unintentional yet beneficial result that only S need be varied is not due to simplification but rather because S is the major contributor to the lines that need to fluctuate in order to obtain a reasonable fit.  In order to reduce the model flux longward of 50 \AA\ to less than $2\sigma$ above the noise, Si and N were depleted to 44\% of cosmic.

\begin{figure} [htbp]
   \centering
   \plotone{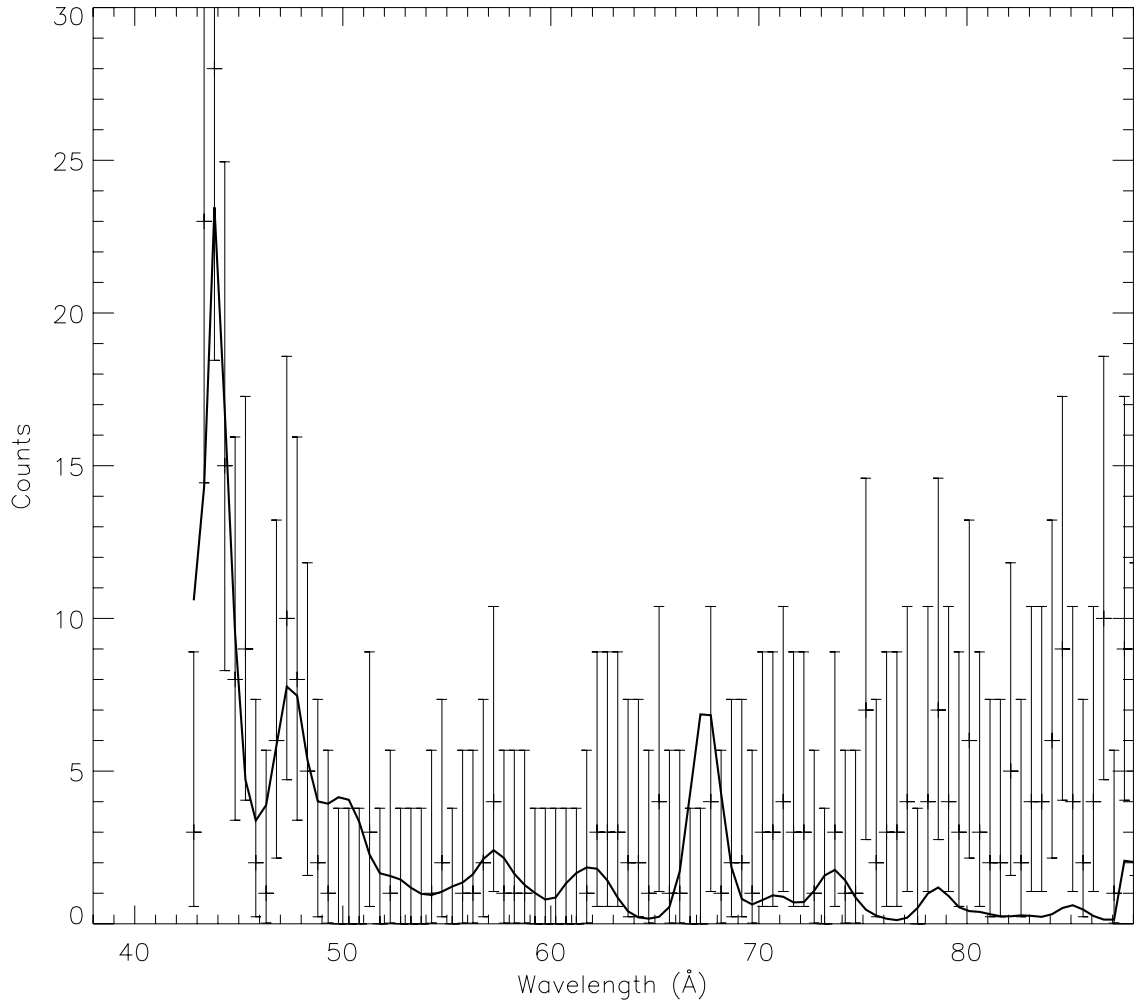}
   \caption{Data are plotted as points with $2\sigma$ errorbars.  The $\log(T)=6.2$ MEKAL equilibrium plasma model is plotted as the solid line.}
   \label{model}
\end{figure}

\begin{figure} [htbp]
   \centering
   \plotone{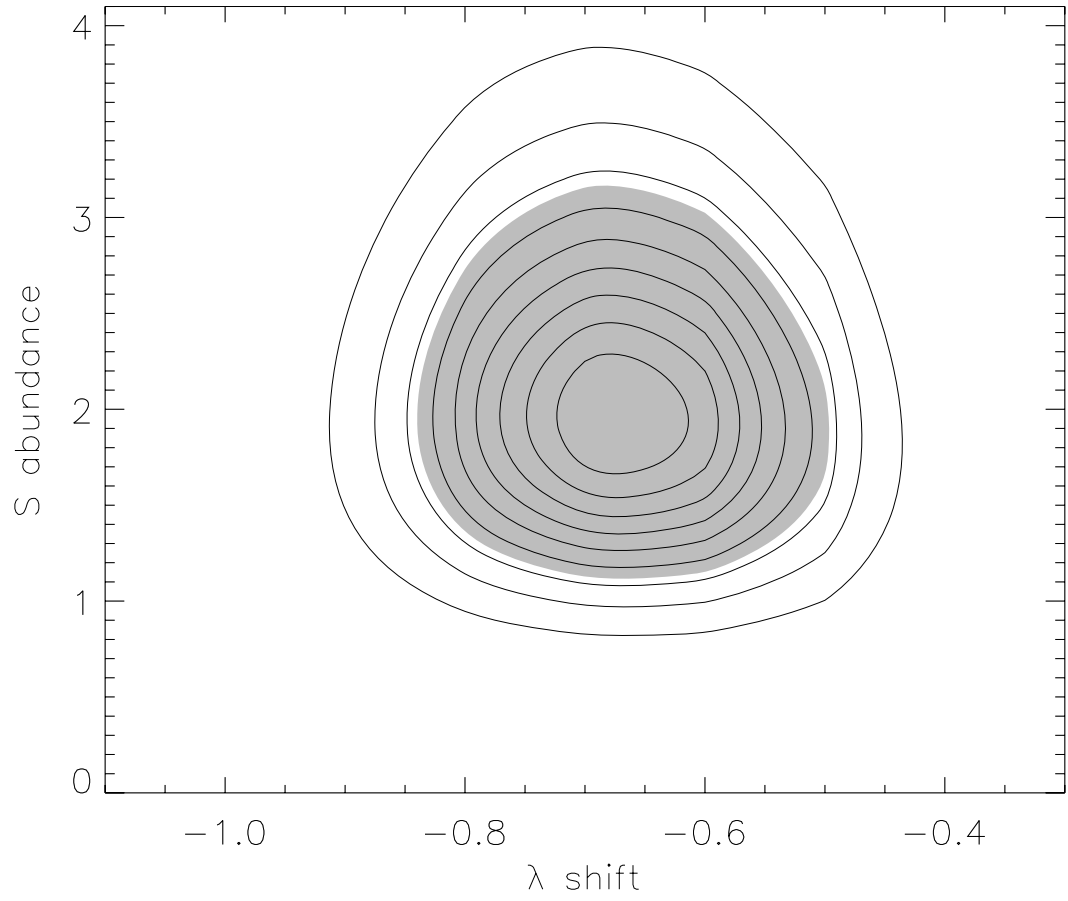}
   \caption{Likelihood contours for the fit parameters are plotted in steps of 10\% of the peak value.  The gray area encompasses the 68\% confidence region.}
   \label{likeli}
\end{figure}

\begin{figure} [htbp]
   \centering
   \plotone{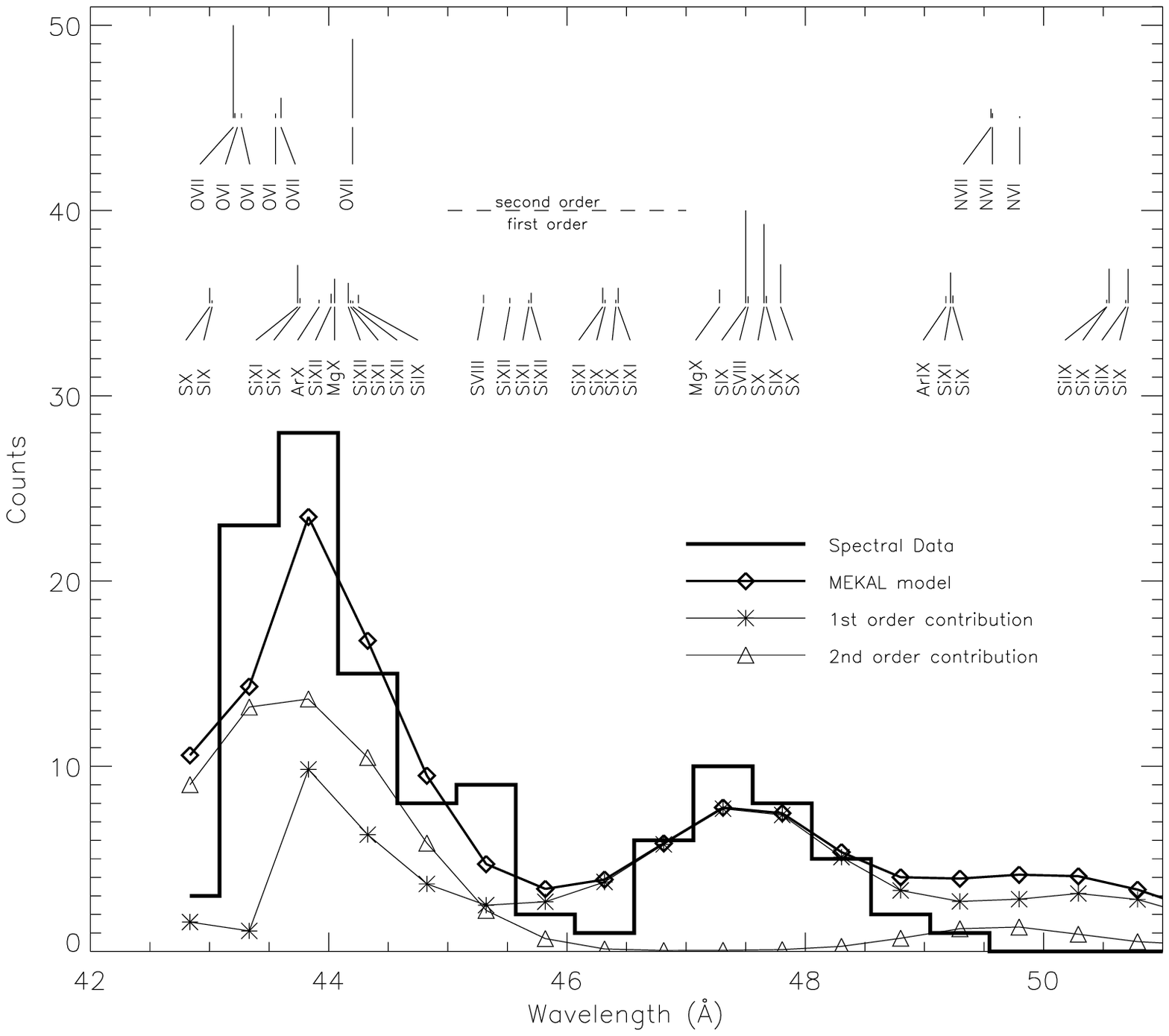}
   \caption{Line identifications for the spectral data.  At the top, the ion names point to vertical lines that depict their relative contributions.  Second order lines are called out above the first order lines.  Spectral data are plotted as the histogram.  The MEKAL plasma fit is plotted as diamonds with first order model contributions as asterisks and second order contributions as diamonds.}
   \label{lines}
\end{figure}

The data and the model (solid curve) are plotted together in figure~\ref{model}.  These data are identical to those shown in figure~\ref{rawdata}, but this time are plotted with 0.9772 single-sided upper and lower limits (0.9544 confidence level) as calculated by \citet{gehrels}, which correspond to $2\sigma$ Gaussian statistics.  The likelihood contours for the fit parameters are given in figure~\ref{likeli}.  The shaded region encompasses 68\% of the normalized likelihood and establishes the 84\% ($1\sigma$ Gaussian) marginalized confidence intervals for the individual parameters, $1.55^{+0.90}_{-0.63}$ for the S abundance and $-0.76^{+0.18}_{-0.17}$ \AA\ for the $\lambda$ shift.  Finally, a closeup of the spectral data along with line identifications are shown in figure~\ref{lines}.  The more prominent data line around 44 \AA\ contains some first order \ion{Mg}{10} and \ion{Si}{9}-\ion{Si}{12}, but most of the flux is in the He-like triplet of \ion{O}{7} in second order.  The other data line around 47-48 \AA\ is dominated by \ion{S}{9} and \ion{S}{10} in first order.  A summary of the major lines is given in table~\ref{tbl1}.  The elemental abundances with respect to cosmic \citep{allen} are C, 1.0; N, 0.44 (max); O, 1.0; Ne, 1.0; Mg, 1.0; Si, 0.44 (max); S, $1.55^{+0.90}_{-0.63}$; Ar, 1.0; Ca, 1.0; Fe, 1.0; Ni, 1.0.  Tabel \ref{tbl1} summarizes the major lines and transitions.




\section{Discussion}\label{disc}
The results of the data analysis reveal a departure from cosmic abundances; S is enriched while Si and N are depleted.  The enrichment of S can be explained by confusion due to multiple temperature components.  The single temperature fit at $kT_e=0.14$ keV is intermediate in comparison to \citet{L99} and \citet{M01}, but consistent with the low temperature component from \citet{M07}.  Therefore, there may be some contamination from a higher temperature component in the CyXESS data.  In this passband, increasing the temperature increases the contribution from second order oxygen while decreasing the contributions of other lines.  Therefore, a high temperature component will only add flux to the line complex at 44 \AA, thus requiring additional flux in the form of a S enhancement in the 47.5 \AA\ complex in order to maintain the spectral shape defined by the fit.  Since there are only 2 features to fit, it is impossible to accurately discern the relative contribution from each plasma.  More spectral resolution is required so that individual lines can be used to define the components. 

As for the second issue, even though some previous observations argue for an equilibrium plasma, especially in the case of the softest X-rays which occur in cloud shocks, nonequilibrium conditions could still be important.  If nonequilibrium is considered, then the ionization state of the gas should be decreased.  In nonequilibrium the electron temperature is higher than what is expected from the ion state of the gas and the gas is underionized.  Therefore, favoring lower ion species in the MEKAL model will test nonequilibrium conditions.  Also, within the lower ion species the ratio of higher energy line flux to lower energy line flux should be increased relative to equilibrium since the electrons have more energy to excite the ions to higher levels than typical in equilibrium.  In the case of Si, the lines that contribute the problematic long wavelength flux occur around 49-50 \AA\ and are \ion{Si}{9}, \ion{Si}{10} and \ion{Si}{11}.  Decreasing the influence of these lines will definitely have a desired effect, but will also require an increase of major \ion{Si}{7} and \ion{Si}{8} lines at 52 \AA, thus reintroducing the problem.  This occurs for other elements as well.  Favoring lower ion species will not explain the depletions, and furthermore, there is no evidence for nonequilibrium conditions contributing to these data.

Another explanation could be depletion into dust, especially in the case of Si depletion.  If dust is a major constituent of the shocked ISM then refractory elements could be contained within the dust and depleted from the gas phase.  However, the dust present in the cloud must be able to survive not only the initial blast wave, but also a reflection shock and sublimation over time in the hot gas (not to mention precursor winds, cloud-cloud interactions, cosmic rays and photodesorption \citep{DraineSalp1,DraineSalp2}).  Given the conditions present in the Cygnus Loop, \citet{DraineSalp1} show that thermal sputtering rates for dust grains in $\log(T)=6.2$ gas are $\sim$0.001$\mu$m every 1000 years.  Therefore, small grains (size $\lesssim$100\AA) will be quickly destroyed with the mass fraction returning to the gas.  In addition, \citet{DraineSalp2} have modelled dust sputtering as a function of blast wave velocity.  Their results show that a blast wave with shock velocity $>300$ km s$^{-1}$ will sputter nearly all graphite, silicate, and iron dust grains up to a size of 0.1 $\mu$m in a cloud with density $n_H=10-100$ cm$^{-3}$.  Calculations for the initial blast wave velocity for the Cygnus Loop vary from 330 km s$^{-1}$ \citep{L02} to 400 km s$^{-1}$ \citep{Ku}.  This also suggests that if a significant fraction of refractory elements are depleted into dust in the dense clouds consituting the Cygnus Loop cavity walls, then the size distribution of dust particles favors large grains.  

Infrared (IR) emission from the Cygnus Loop has been observed and \citet{arendt} show that it can be explained by dust emission.  Using \textit{IRAS} observations at 12 $\mu$m, 25 $\mu$m, 60 $\mu$m, and 100 $\mu$m, they find that there are two infrared (IR) components, one that correlates well with X-ray emission and another that correlates with the optical emission.  A lack of observed emission in the 12 $\mu$m and 25 $\mu$m bands suggests an underabundance of small grains.  This is supported by their models of the X-ray/IR correlated gas, which favor a minimum grain size of $\sim$150 \AA\, consistent with our estimates.  However, the models used to fit the broadband observations assume all emission is due to thermal dust emission.  The authors state that line emission could theoretically contribute a significant fraction to the observed IR emission.  Therefore, constraining the dust fraction is impossible without higher resolution IR spectra, which should become available with upcoming Spitzer observations.

Depletion into dust cannot be ruled out and is supported by our data, especially in the case of silicate grains.  We see no evidence for graphite grains because there are no important C lines in the wavelength range of our spectrum.  The N and Si depletions necessary for our fit are due to a line complex of second order \ion{N}{7} and first order \ion{Si}{10} and \ion{Si}{11}.  Therefore, given this confusion only an upper limit to the Si abundance can be applied suggesting a large depletion of at least 56\% of the Si into grains.

\section{Summary}\label{sum}
The CyXESS payload was designed to observe the soft X-ray flux of the Cygnus Loop supernova remnant.  The design consisted of a wire grid collimator that focused the light onto an array of gratings in the off-plane mount which ultimately dispersed the spectrum onto large format GEM detectors.  The payload was launched on November 20th, 2006 from White Sands Missile Range and collected 345 seconds of data.  Data reduction decreased the amount of usable data to 65 seconds during which the instrument detected flux between 43-49.5 \AA\ (250-288 eV) in two prominent features.  The first feature near 45 \AA\ is dominated by the He-like triplet of \ion{O}{7} in second order with contributions from \ion{Mg}{10} and \ion{Si}{9}-\ion{Si}{12} in first order, while the second feature near 47.5 \AA\ is first order \ion{S}{9} and \ion{S}{10}.  Fits to the spectra give an equilibrium plasma at $kT_e=0.14$ keV ($\log(T)=6.2$) and near cosmic abundances for most elements.  Even though the most likely fit to the CyXESS data contains only one temperature, the lack of spectral range and resolution do not allow determination of multiple components.  An observed depletion in Si supports the presence of silicate grains but higher X-ray and IR resolution are necessary to accurately constrain dust models.

Our data were constrained to a small portion of the soft X-rays, but show a wealth of lines present.  In this band we find no evidence to support nonequilibrium conditions.  Therefore, our results are consistent with previous observations and show that the soft X-ray spectrum of the Cygnus Loop, which is dominated by interactions between the initial blast wave with the walls of a precursor formed cavity, can be described by an equilibrium plasma.

\acknowledgments
The authors would like to acknowledge NASA grants NNG04WC02G and NGT5-50397 for support of this work.

\clearpage

\clearpage

\clearpage

\clearpage

\clearpage

\clearpage

\clearpage

\clearpage

\clearpage

\clearpage

\begin{table}[tbp]
\centering
\caption{Major spectral lines detected by CyXESS.\label{tbl1}}
\begin{tabular}{cclc}
\hline \hline
Ion	&	Wavelength (\AA)	&	Transition	&	Relative strength	\\ \hline
\ion{O}{7}	&	21.600	&	1s$^2$ - 1s 2p [R]	&	1.000	\\
\ion{O}{7}	&	21.800	&	1s$^2$ - 1s 2p [I]	&	0.216	\\
\ion{O}{7}	&	22.100	&	1s$^2$ - 1s 2s [F]	&	0.852	\\
\ion{Si}{11}	&	43.740	&	2s$^2$ - 2s 3p	&	0.182	\\
\ion{Mg}{10}	&	44.050	&	2s - 4p	&	0.116	\\
\ion{Si}{12}	&	44.165	&	2p - 3d	&	0.096	\\
\ion{S}{9}	&	47.500	&	2p$^4$ - 2p$^3$ 3d	&	0.444	\\
\ion{S}{10}	&	47.654	&	2p$^3$ - 2p$^2$ 3s	&	0.378	\\
\ion{S}{10}	&	47.793	&	2p$^3$ - 2p$^2$ 3s	&	0.186	\\ \hline
\end{tabular}
\end{table}

\end{document}